\documentclass[a4paper]{revtex4}
\usepackage{graphicx}
\usepackage{fancyhdr}
\usepackage{amsmath}
\usepackage{color}

\begin{document}

\title{Dark matter in the nonabelian hidden gauge theory}

%

\author{Nodoka~Yamanaka$^1$, Sho~Fujibayashi$^2$, Shinya~Gongyo$^3$, Hideaki~Iida$^2$}
\affiliation{
$^1$iTHES Research Group, RIKEN, 
Wako, Saitama 351-0198, Japan,\\
$^2$Department of Physics, Graduate School of Science,
Kyoto University, 
Kitashirakawa-oiwake, Sakyo, Kyoto 606-8502, Japan,\\
$^3$Yukawa Institute for Theoretical Physics, Kyoto University, Kyoto 606-8502, Japan
}

\begin{abstract}
We discuss the dark matter in the hidden gauge theory.
We propose a scenario where the mini-inflation dilutes the dark matter density.
This scenario is consistent with the current baryon number asymmetry.

\end{abstract}

\maketitle

\thispagestyle{fancy}


\section{Introduction}

Recent observations suggest that 27\% of the energy of our Universe is composed of dark matter \cite{planck}.
As a natural candidate of theories explaining the dark matter, we have the hidden gauge theory (HGT) \cite{hgt,secluded,compositedm,chiral}.
In this scenario, the lightest particles are hadrons (pions or glueballs).
The scale of the HGT is controlled by the color and flavor numbers, and this fact makes the conception of the HGT quite natural, avoiding serious hierarchical problems.
If the HGT is unified at the grand unification scale with the gauge interactions of the standard model (SM), the scenario is even more natural.
In this talk, we discuss the HGT and propose a scenario where the Higgs mini-inflation dilutes the dark matter.

\section{Hidden gauge theory}

The lagrangian of the HGT is given by
\begin{equation}
{\cal L} =
-\frac{1}{4} F^{\mu \nu}_a F_{\mu \nu}^a + \sum^{N_f}_q \bar q (D\hspace{-.65em}/\, -m_q ) q 
,
\end{equation}
where $q$ is the fermion of the hidden sector, with $N_f$ flavors.

In the HGT, the dark matter particle is a glueball if there are no quarks lighter than the scale parameter of the HGT $\Lambda_{\rm DM}$.
In the opposite case, it is a pion.
From dimensional analysis, the masses of the glueballs and pions are respectively given by
\begin{eqnarray}
m_\phi
&\sim &
\Lambda_{\rm DM} ,
\label{eq:glueballmass}
\\
m_\pi
&= &
\frac{1}{f_{\rm DM}} \sqrt { m_q \langle 0 | \bar q q |0 \rangle }
\sim
\sqrt{ m_q \Lambda_{\rm DM}}
.
\label{eq:pionmass}
\end{eqnarray}
Here $f_{\rm DM}$ and $\langle 0 | \bar q q |0 \rangle$ are, the pion decay constant and the chiral condensate, respectively.
They scale as $ \Lambda_{\rm DM}$ and $ \Lambda_{\rm DM}^3$, respectively.

\section{Phenomenology}

The dark matter relic density is given by the freeze-out temperature $T_{\rm FO}$, where the dark matter particles are considered to become nonrelativistic.
The number density of the relic dark matter hadrons is given by
\begin{equation}
n_{\rm DM} (T_{\rm FO})
=
g_{\rm DM} (T_{\rm FO}) \Bigl( \frac{m_{\rm DM} T_{\rm FO}}{2\pi } \Bigr)^{\frac{3}{2}} e^{-\frac{m_{\rm DM}}{T_{\rm FO}}} 
\approx 
O(m_{\rm DM}^3 )
,
\label{eq:boltzmann}
\end{equation}
where $m_{\rm DM}$ is the mass of the lightest dark matter hadron.
The effective degree of freedom was considered to be $g_{\rm DM} = O(1)$, and $T_{\rm FO} \sim m_{\rm DM}$.
The dark matter density can be extrapolated from the current density to the freeze-out temperature as
\begin{equation}
n_{\rm DM} (T_{\rm FO})
=
\frac{\rho_{{\rm m}0}}{m_{\rm DM}}
\frac{
T_{\rm FO}^3 
}{
\xi^3 T_{\rm eq}^3 a_{\rm eq}^3
}
,
\label{eq:entropyconservation}
\end{equation}
where $a_{\rm eq}$ is the scale factor at the time of matter-radiation equality.
The parameter $\xi$ is the entropy ratio between the SM sector and the hidden sector.
By equating Eqs. (\ref{eq:boltzmann}) and (\ref{eq:entropyconservation}) with $T_{\rm FO}\approx m_{\rm DM}$, we obtain
\begin{equation}
m_{\rm DM}
\approx 
T_{\rm FO} 
= 
\xi \times
O(10^{-8}) {\rm GeV}
.
\label{eq:naivetfo}
\end{equation}
If $\xi$ is not shifted, the dark matter will remain as a radiation until the recombination.
This fact however contradicts with the structure formation ($T_{\rm FO} > O(10)$ eV), the bigbang nucleosynthesis ($T_{\rm FO} > O(1)$ MeV), and the scale of the self-interaction of the dark matter ($T_{\rm FO} = {\rm MeV} \sim {\rm TeV}$) \cite{dmself-interaction}.

\begin{figure}[ht]
\centering
\includegraphics[width=80mm]{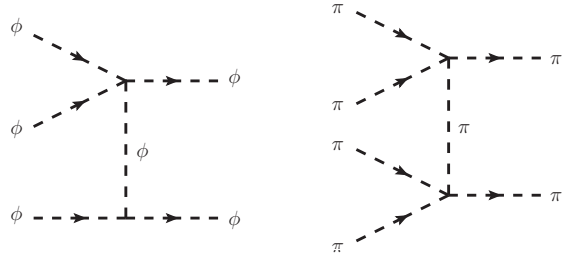}
\caption{
Examples of diagrams contributing to the decay of the halo.
} \label{fig:annihilation}
\end{figure}

The stability of the dark matter halo provides us another constraint.
If there is no baryon number asymmetry in the hidden sector, the dark matter particles may reduce their number through the processes depicted in Fig. \ref{fig:annihilation}.
This number reducing annihilation shrinks the dark matter halo by emitting relativistic products, and therefore contributes to the decay of the halo.
By using the dimensional analysis, the decay rate of the halo can be estimated as
\begin{eqnarray}
\Gamma_3
&\sim &
\frac{\rho_0^3 R^3}{\Lambda_{\rm DM}^8}
,
\label{eq:Gamma_3}
\\
\Gamma_4
&\sim &
\frac{\rho_0^4 R^3}{m_q^2 \Lambda_{\rm DM}^{10}}
,
\label{eq:Gamma_4}
\end{eqnarray}
where we have assumed a constant distribution of the halo with $\rho_0 \sim 0.3\, {\rm GeV}/{\rm cm}^3$
with the radius $R \sim 20$ kpc.

The stability of the dark matter halo requires $\Gamma < H_0$ where $H_0 \equiv \frac{\dot{a}_{t=t_0}}{a_0}=67.80 \pm 0.77 \,{\rm km\ s}^{-1} {\rm Mpc}^{-1}$.
We thus have
\begin{eqnarray}
\Lambda_{\rm DM}
&>&
10^4 \, {\rm GeV},
\label{eq:Gamma_3<H_0}
\\
\Lambda_{\rm DM}
&>&
\Bigl( \frac{\Lambda_{\rm DM}}{m_q} \Bigr)^{\frac{1}{6}} \times 10^{-1} \, {\rm GeV}
.
\label{eq:Gamma_4<H_0}
\end{eqnarray}
The first inequality corresponds to the case of the glueball dark matter, and the second one to the pion dark  matter.
These constraints are consistent with the scale of the dark matter self-interaction.
The HGT scenario thus requires $\xi \sim 10^{12}$ for the glueball dark matter, and $\xi \sim 10^7$ for the pion dark matter [we have assumed $( \Lambda_{\rm DM} / m_q)^{1/6} \sim O(1)$].

From the above constraints, the increase of the entropy of the visible sector is required.
There are three ways to realize it.
The first possibility is to make the asymmetry of the entropy at the grand unification temperature.
The second is to connect the dark and visible sectors with some light mediators.
The last case is the generation of the entropy in the visible sector with inflation.

The first possibility may be realized for instance through the asymmetric decay of the inflaton.
In this work, we do not consider such case.

\begin{figure}[ht]
\centering
\includegraphics[width=60mm,angle=270]{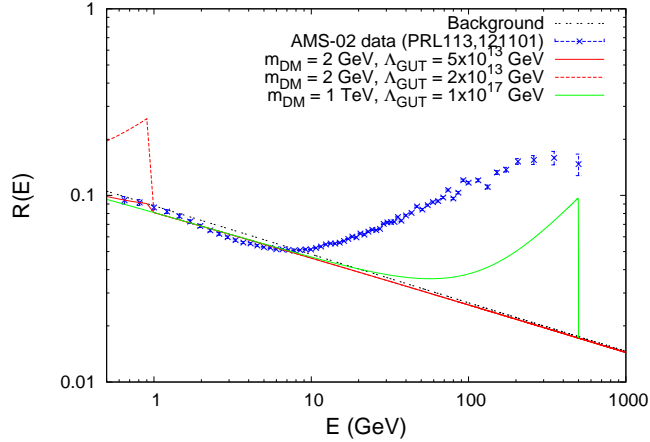}
\caption{
The positron fraction calculated with several dark matter and mediator masses.} \label{fig:positron_fraction}
\end{figure}

In the HGT, the second scenario is ruled out from the indirect detection experiments of dark matter.
Let us consider the electron/positron pair production from the decay of the DM hadron with the life-time
\begin{equation}
\tau_{\rm DM} 
\sim
\frac{\Lambda_{\rm GUT}^4}{m_{\rm DM}^5}
,
\end{equation}
where $\Lambda_{\rm GUT}$ is the mass scale of the mediator.
The positron fraction can be calculated by considering the diffusion following Ref. \cite{baltz} and using the background contribution of Ref. \cite{ibe}.
The result is shown in Fig. \ref{fig:positron_fraction}.
From the experimental data of AMS-02 \cite{ams-02}, we have
\begin{equation}
\Lambda_{\rm GUT} > O(10^{13}) {\rm GeV}
,
\end{equation}
for $m_{\rm DM} \sim 1$ GeV.
This result suggests that the mediator between the hidden and visible sectors is not light, and the second scenario is ruled out.
We are therefore left with the third possibility, the increase of the entropy of the visible sector through inflation.

\section{Higgs mini-inflation}

\begin{figure}[ht]
\centering
\includegraphics[width=160mm]{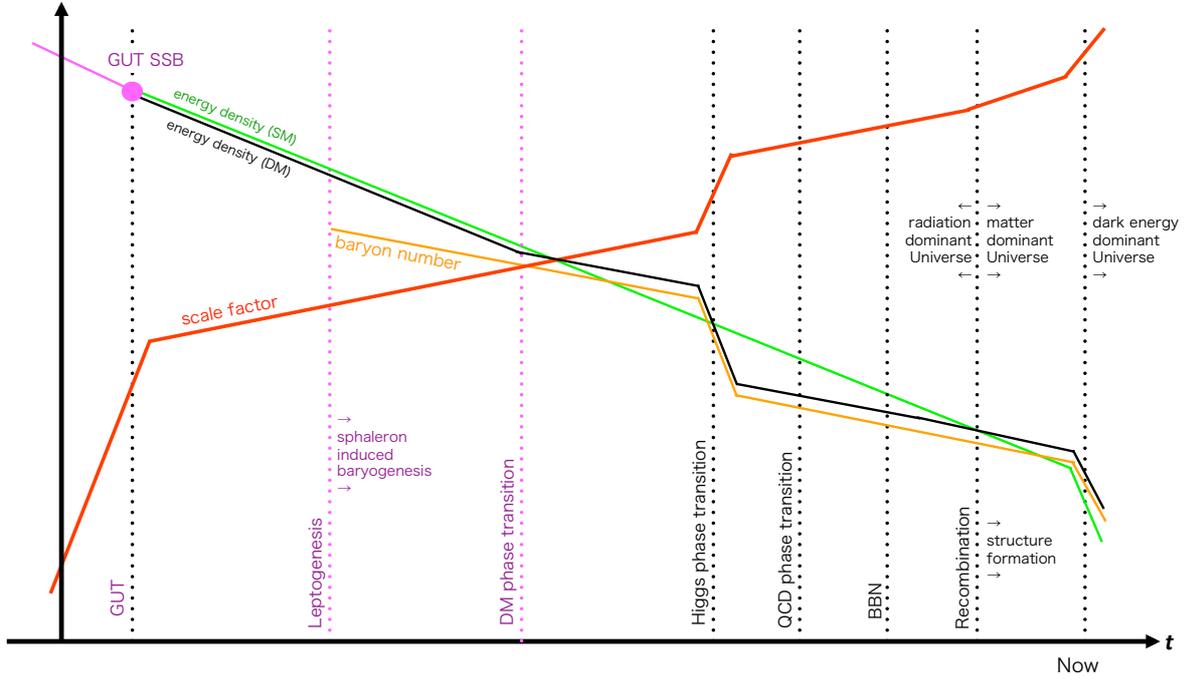}
\caption{
Example of the thermal history with Higgs mini-inflation.} \label{fig:thermal_history}
\end{figure}

We now propose a scenario where the Higgs potential beyond SM induces some mini-inflations \cite{higgsinflation}.
The Higgs mini-inflation dilutes the DM density, and the Higgs potential coupled to the SM will generate large entropy due to the reheating.
The above scenario has advantages, such as the dilution of the baryon number asymmetry.
Currently, the observed  baryon-to-photon ratio is $\frac{n_B}{n_\gamma} = 6 \times 10^{-10} $.
In the most natural case, $\frac{n_B}{n_\gamma}$ should be $O(1\mathchar`-0.01)$, and a dilution of $\sim O(10^{-7})$ is required.
A sudden expansion of the scale factor of $(a_{\rm f} / a_{\rm i} )  \sim 10^2\mathchar`-10^3$ due to the mini-inflation yields an enhancement of the entropy of order of $\xi \sim 10^6 \mathchar`-10^9$, making a consistent dilution of the baryon-to-photon ratio and the dark matter density with $m_{\rm DM} \sim $ GeV.
The Higgs mini-inflation is thus a good candidate for realizing this scenario.
The thermal history is shown in Fig. \ref{fig:thermal_history}.

We now give the experimental observables to probe our scenario.
The first observable is the stochastic gravitational wave background, which is radiated at the first order phase transition \cite{grojean}.
It may determine the critical temperature of the HGT, if it is of first order.
Those signals will be observed with future gravitational wave observation experiments \cite{ska}.

The second observable is the precision test of the Higgs potential beyond SM.
The Higgs potential may be accurately studied using next-generation linear colliders.
There the possibility of the mini-inflation may be examined.

The third observable is the indirect detection of cosmic rays.
The cosmic ray spectrum probes the scale of the mediator and also the scale of the grand unification $\Lambda_{\rm GUT}$.
We are waiting for the data of the spectrum above TeV.

The final observable is the density profile of the dark matter halo.
By comparing with the result of simulations, the self-interaction of the dark matter particles may be unveiled.

\section{Conclusion}

In this work we have studied the HGT as a natural candidate of theories explaining the dark matter.
From the phenomenological analysis, the scale of the HGT is bound by many constraints such as the structure formation.
We have pointed that the stability of the dark matter halo gives a strong constraint on this scale ($>$ 1 GeV).
Within the above constraint, the entropy of the visible sector must be increased relatively to the hidden sector.
The transfer of the entropy of the hidden sector to the visible sector at the freeze-out is ruled out due to the constraint from the cosmic ray spectrum.

To resolve this problem, we have proposed the Higgs mini-inflation to dilute the dark matter.
The dilution due to the mini-inflation is also consistent with the current baryon-to-entropy ratio.
We have also proposed four possible observables to probe this scenario.


\bigskip 

\end{document}